\documentclass[12pt]{article}

\usepackage{sbc-template}
\usepackage{graphicx,url}
\usepackage[utf8]{inputenc}

\usepackage{hyperref} 

\usepackage{multirow}
\usepackage{array}

\usepackage{todonotes}

\title{Audio MFCC-gram Transformers for respiratory insufficiency detection in COVID-19\\~\\}

\author{Marcelo Matheus Gauy\inst{1}, Marcelo Finger\inst{1}}

\address{Instituto de Matem\'atica e Estat\'istica -- Universidade de S\~{a}o Paulo (USP)\\}

\begin{document}

\maketitle
\begin{abstract}
This work explores speech as a biomarker and investigates the detection of respiratory insufficiency (RI) by analyzing speech samples. 
Previous work~\cite{spira2021} 
constructed a dataset of respiratory insufficiency COVID-19 patient utterances and analyzed it by means of a convolutional neural network achieving an accuracy of $87.04\%$, validating the hypothesis that one can detect RI through speech. 
Here, we study how Transformer neural network architectures can improve the performance on RI detection. 
This approach enables construction of an acoustic model. By choosing the correct pretraining technique, we generate a self-supervised acoustic model, leading to improved performance ($96.53\%$) of Transformers for RI detection.



\end{abstract}

\section{Introduction}\label{sec:intro}

COVID-19 is the cause of a major pandemic that threatens to collapse the healthcare systems in many regions of the world. Respiratory insuficiency (RI) is one of COVID-19 symptoms, which often requires hospitalization and is aggravated by a common COVID-19 condition called \textit{silent hipoxia}, low blood oxygen concentration without breath shortness~\cite{tobin2020covid}. This work aims to help deal with the COVID-19 pandemic by providing an automated system, based on deep learning techniques, capable of detecting RI in COVID-19 patients. Such an automated system could, for example, support cellphone-based patient triage procedures alleviating the burden on health personnel.

We explore the view of \textit{speech as a biomarker}, by building upon a recently shown fact: it is possible to detect respiratory insufficiency through analyzing spoken utterances in real-life conditions (typically a moderately large sentence). This hypothesis has been previously verified~\cite{spira2021} by using a CNN-based deep neural network. This CNN received a moderately large sentence spoken in real life conditions and had to predict whether it came from a patient with RI or from the control group. In this work, we aim to further analyze that hypothesis by studying other network architectures (namely, Transformers~\cite{vaswani2017attention}), in an attempt to improve the results previously obtained in~\cite{spira2021}, with a view of extending it in the future to RI originated from other causes, such as influenza, heart disease or mental illness.


In this work we find that Transformers can be used for detecting respiratory insufficiency with an accuracy of $96.38\%$ up from $87.04\%$ in~\cite{spira2021}. To reach that level of performance, we feed the Transformers with a sequence of Mel Frequency Cepstral Coefficients (MFCC) obtained from the patients' audios (henceforth called MFCC-gram Transformers). Like CNN-based detection  from~\cite{spira2021}, the Transformer performance drops significantly (to $82.87\%$) if we feed it standard spectrogram coefficients (called Spectrogram Transformers after~\cite{gong2021ast}).

%
The Transformers~\cite{vaswani2017attention} were shown to be very effective when divided in two parts~\cite{devlin2018bert}.  The \textit{pretraining phase}  generates a language-based acoustic model with unsupervised (or self-supervised) learning by optimizing a generic language prediction task with a large amount of generic data.  Then, the  acoustic model undergoes a task-specific \textit{refinement phase} in which both the acoustic model and additional task-specific neural modules are trained on smaller-size application data. A \textit{baseline transformer} is one in which pretraining is a random assignment of weights.

Here, we find that MFCC-gram Transformers benefit from being pretrained with large quantities of spoken Brazilian Portuguese audios, which is later refined for the target task of detecting respiratory insufficiency. For pretraining, we explore three known techniques from the literature~\cite{liu2020mockingjay, liu2020tera} and find that they generally lead to some performance improvement over baseline transformers. Performance reaches $96.53\%$ using the best of the available techniques. 

\section{Related Work}

In addition to~\cite{spira2021} there have been other works~\cite{pinkas2020sars, laguarta2020covid} which study COVID-19 with deep learning using voice related data. \cite{pinkas2020sars} attempt to detect SARS-COV-2 (the virus that causes COVID-19) from voice audio data, while this work and~\cite{spira2021} attempt to detect RI. Furthermore, there have been previous works which support the view of speech as a biomarker~\cite{botelho2019speech, nevler2019validated, robin2020evaluation}.

Transformers were designed for NLP~\cite{vaswani2017attention, devlin2018bert}, and were also later used in audio processing tasks~\cite{liu2020mockingjay, liu2020tera, schneider2019wav2vec, baevski2020wav2vec, baevski2019vq, song2019speech}. In Mockingjay and Tera~\cite{liu2020mockingjay, liu2020tera}, it was used in phoneme classification and speaker recognition tasks. There it was shown that variants of the Cloze task~\cite{taylor1953cloze, devlin2018bert} for audio could be used for unsupervised pretraining of Transformers. In Wav2Vec and its variants~\cite{schneider2019wav2vec, baevski2020wav2vec, baevski2019vq}, a contrastive loss is used to enable unsupervised pretraining, which is later finetuned to speech and phoneme recognition tasks. In Speech-XLNet~\cite{song2019speech}, a speech based version of the XLNet~\cite{yang2019xlnet} was proposed. The XLNet is a network that maximizes the expected log likelihood of a sequence of words with respect to all possible autoregressive factorization orders.

\section{Methodology}

\subsection{Datasets}

For the task of respiratory insufficiency detection, the data used in the refinement phase is the same one used in~\cite{spira2021}. There, COVID patient utterances were collected by medical students at COVID wards from patients with blood oxygenation level below $92\%$, as an indication of RI. Control data was collected by voice donations over the internet without any access to blood oxygenation measurements and were therefore assumed healthy. As COVID wards are noisy locations, an extra collection was made consisting of samples of pure background noise (no voice). This is a crucial step in preventing the network to overfit to the background noise differences in data collection.

The gathered audios contained $3$ utterances:
\begin{itemize}
\item A long sentence with $31$ syllables. It was designed by linguists to be long enough to have reading pauses while being simple for even low literacy donors to speak.
\item A widely known nursery rhyme for readers with reading impediments.
\item A well known song along the lines of 'Happy birthday to you'.
\end{itemize}

As suggested in~\cite{spira2021}, we select only audios from the first utterance and sample balance the dataset by class and sex. The presence of ward background noise in the patient audios is treated in a similar way: we insert noise to the control group as that is easier than removing it from the patients' signal. This prevents that we eliminate from the signal, audio that is relevant to the network's classification.

We employ the same division in training, validation and test as done in~\cite{spira2021}. The best signal-noise ratio audios are included in the test set. The second best audios are in the validation set. This is done to detect training overfitting. Table~\ref{table:dataset} contains information on the number of audio files for each class.

\begin{table}[htb]
\begin{tabular}{|m{4em}|m{2.5em}|m{2.5em}|m{4em}|m{2.5em}|m{2.5em}|m{4em}|m{5em}|}
\hline
\multirow{2}{*}{Sets} & \multicolumn{3}{c|}{Control} & \multicolumn{3}{c|}{Patients} & \multirow{2}{*}{Total Audios} \\ \cline{2-7}
                  &   Male        & Female   & Mean duration(s)      &    Male       &    Female    & Mean duration(s)   &                   \\ \hline
     Training             &    $59$       &     $84$    & 8.15 &     $83$      &     $66$    & 13.18 &    $292$               \\ \hline
    Validation              &   $8$        &    $8$    & 7.75  &     $8$      &  $8$   &  10.78    &   $32$                \\ \hline
    Test              &     $22$      &     $26$   & 7.77  &     $28$      &     $32$  &  9.43  & $108$                  \\ \hline
\end{tabular}
\caption{Filtered dataset information.}\label{table:dataset}
\end{table}

For the pretraining phase, we use datasets containing Brazilian Portuguese speech. These datasets are NURC-Recife~\cite{oliviera2016nurc}, ALIP~\cite{gonccalves2019projeto}, C-Oral Brasil~\cite{raso2012c} and SP2010~\cite{mendes2013projeto}. Together, they contain more than $200$ hours of Brazilian Portuguese speech.

\subsection{Preprocessing}

As we face similar audio processing issues as~\cite{spira2021}, we employ similar preprocessing steps. In the dataset, the majority of audios were sampled at $48kHz$. We preprocess the files using Torchaudio $0.9.0$. We extract either the mel-spectrogram (for Spectrogram Transformers) or the MFCCs of the audios with default Torchaudio parameters and retain $128$ coefficients. Torchaudio, by default, employs a Fast Fourier Transform~\cite{brigham1967fast} with a $400ms$ window and hop length $200$.

As the dataset has an inherent imbalance in the audio lengths from patients and control we do not use the full audios of the first utterance. Instead, we break each audio into $4$ seconds chunks, with a windowing of $1$ second steps. Such a windowing method was observed in~\cite{spira2021} to be more effective than, for example, padding the audios with zeros to make all the audios have the same length. The windowing technique solves the problem of the imbalance between audio lengths and guarantees the network will not pay too much attention to the audio lengths and instead focuses on the content. The windowing technique also serves as a kind of data augmentation as, for example, an audio with $8$ seconds becomes $5$ audios with $4$ seconds. We observe that the windowing should be done before the spectrogram or MFCCs feature extraction.

\subsection{Noise insertion}

The noise in COVID wards is a serious bias source. This can be seen in our experiments and in the original work with the dataset by~\cite{spira2021}. One potential way of dealing with this bias source is to filter the noise and eliminate it. However, this has the risk that we eliminate important low-energy information from the data, information which would have been useful in detecting whether a patient had RI. Moreover, eliminating the noise could also create extra biases, as different procedures for eliminating patient and control noises would be required. Thus, instead of eliminating the noise, we consider it much easier to insert the noise present in the COVID wards into all the audio samples.

The original dataset contained $16$ samples of $1$ minute each containing just the background noise present in COVID-19 wards. These noise samples are added to all the training, validation and test audios, similarly to what was done in~\cite{spira2021}. We experiment with the amount of noise we add to each of the audio files. During training, audio samples are injected with one or more noise samples. These are selected randomly from the pool of noise samples each time an audio is used for training. The starting point of each noise sample is also selected randomly. Lastly, a factor to change the intensity of the sample is drawn. This factor is limited by a maximum amplitude value which depends on the patient audio noises. This process is similar to the one in~\cite{spira2021} and the goal is inserting noise as similar to the pre-existing noise as possible.

\subsection{Transformers}


We consider two types of Transformers: MFCC-gram Transformers and Spectrogram Transformers. They are equivalent except in the data features that are fed to them: MFCC-gram Transformers receives MFCC audio features and Spectrogram Transformers receive mel spectrogram audio features. Our Transformers are equivalent to the Transformer Encoder units described in~\cite{vaswani2017attention}. Namely, we use a multi-layer Transformer encoder ($3$ layers) with multi-head self-attention. Each encoder layer has two sub-layers, the first being a multi-head self-attention network and the second being a fully connected feed-forward layer. Each sub-layer has a residual connection followed by layer normalization~\cite{ba2016layer}. Every encoder layer and sub-layers produce outputs of dimension $512$. In addition to the attention sub-layers, each encoder layer contains a fully connected feed-forward network with an inner layer of dimension $2048$.

In order to generate the sequence of tokens that is sent to the Transformers the MFCC and/or Spectrogram is split into its frames. Each frame of the MFCC or spectrogram corresponds to one token fed to the sequence. We also attempted joining multiple frames into one token but this typically produced worse results than the one to one framework. We use sinusoidal positional encoding~\cite{vaswani2017attention, liu2020mockingjay, pham2019very} to make our model position aware. As suggested by~\cite{liu2020mockingjay}, each frame is first projected linearly to a hidden state of dimension $512$.

Our Transformers are trained in two phases: pretraining and refinement. In the pretraining phase, we leverage the unsupervised training techniques described in Section~\ref{subsec:unsupervisedpretraining} to build an acoustic model over generic audio data. In the refinement phase, the pretrained Transformers is refined over COVID related audio data. For some experiments, we bypass the pretraining phase by initializing the Transformers with a random assignment of weights and refining that over the COVID data. This is done to get a baseline performance and we call these Transformers the baseline Transformers. We will name our Transformers types baseline MFCC-gram Transformers and baseline Spectrogram Transformers when we consider Transformers which bypass the pretraining phase.

Our code is based on the guide ``The annotated Transformer''\footnote{\url{http://nlp.seas.harvard.edu/2018/04/03/attention.html}}. While our Transformers are small in comparison to the ones used, e.g. in BERT~\cite{devlin2018bert}, the amount of available data for respiratory insufficiency detection is also rather small so we do not expect that larger Transformers would yield significantly improved results. Once more data is available, it is recommended to also increase our Transformers. Our code can be found in this~\href{https://github.com/marcelomatheusgauy/Audio_mfcc_gram_transformers}{GitHub} repository.

\subsection{Unsupervised pretraining: acoustic model construction}
\label{subsec:unsupervisedpretraining}

We describe three techniques to pretrain acoustic models in a self-supervised way. They are based off Masked acoustic modelling~\cite{liu2020mockingjay}. This erases a fraction of the input and tries to reconstruct the erased parts from the remaining frames. They are bidirectional methods and the reconstruction depends on both left and right contexts.

\textbf{Time Alteration}: also called Masked acoustic modelling~\cite{liu2020mockingjay}. Start by selecting frames up to $15\%$ of the input\footnote{More precisely, we select a fraction of the frames in chunks of a certain size so that the total number of frames masked amounts to $15\%$. In the experiments, the chunk size was $7$.}, 1) mask them all to zero $80\%$ of the time, 2) replace all with a random frame $10\%$ of the time or 3) leave the frames be in the remaining $10\%$ of the time. The goal of this process (as opposed to always masking the frames) is to alleviate the mismatch between training and inference.

\textbf{Channel Alteration}: this techinique is from~\cite{liu2020tera}. Randomly mask a block of consecutive quefrency channels to zero for all time steps of the input sequence. First, the width $W_C$ of the block is selected uniformly from $\{0,1,\ldots, W\}$ where $W$ is a $10\%$ fraction of the total number of channels. Second, sample a channel index $I_C$ from $\{0,1,\ldots, H-W_C-1\}$ where $H$ is the total number of channels in the input. Then, channels from $I_C$ to $I_C+W_C-1$ are masked to zero. Observe that (as with time alteration), a fraction of the time none of the channels will be masked.

\textbf{Noise Alteration} this technique is from~\cite{liu2020tera}. Apply sampled Gaussian noise to change the magnitude of the inputs with a probability of $10\%$. For that end, we sample a random magnitude matrix with the same size as the input. Each element in the matrix is sampled from a normal distribution with mean zero and $0.2$ variance. The matrix is then added to the real input frames.

\section{Results and Discussion}

Here we show the results obtained by the two experiments performed: the first where we compare baseline MFCC-gram Transformers, baseline Spectrogram Transformers and the CNN from~\cite{spira2021}, and the second where we try different unsupervised pretraining techniques to improve baseline Transformers by building an acoustic model.

First, we note that when no ward noise is added to either the patient or control files, baseline MFCC-gram Transformers performs very well ($98.89\pm 0.38$) in the test files. However, this performance drops dramatically (to $70.07\pm 3.15$) if we add noise to the test files and this is a strong sign the model is biased by the noise. This bias is less extreme than what was observed at the MFCC-gram CNN in~\cite{spira2021} but is still present. Therefore, in our experiments, noise is added to the training and test files.

\begin{figure}[htb]
\includegraphics[width=\textwidth]{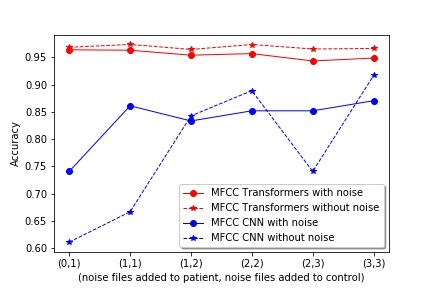}
\caption{This has the same data as Table~\ref{table:comparison}. The y axis shows accuracy and the x axis shows the number of noise files added to patient and control files.}\label{figure:comparison}
\end{figure}

In the first experiment, we consider baseline Transformers and bypass the pretraining phase. We vary the amount of ward noise we add to the training and test files. We add to the audio files between $0$ and $3$ noise files, including either the same amount of noise files to the patient and control audio files or one more file to the control files. This is comparable to the Experiments $3.x$ from~\cite{spira2021} and we can directly compare baseline MFCC-gram Transformers, baseline Spectrogram Transformers with the CNN from~\cite{spira2021}. We perform each experiment for $20$ epochs and repeat the experiments $10$ times. The batch size is set to $16$. The results are in Table~\ref{table:comparison}. We show both the performance when including noise as well as the performance without including noise in the test samples. Figure~\ref{figure:comparison} shows the same data as Table~\ref{table:comparison}.

\begin{table}[htb]
\begin{tabular}{|m{6em}|m{3em}|m{3em}|m{10em}|m{10em}|}
\hline
\multirow{2}{*}{Model} & \multicolumn{2}{c|}{Noise Samples} & \multirow{2}{10em}{Accuracy (with noise in test samples)} & \multirow{2}{10em}{Accuracy (without noise in test samples)} \\ \cline{2-3}

 &    Patient      &    Control       &                   &                   \\ \hline
\multirow{6}{6em}{Baseline MFCC-gram Transformers} &  0      &   1        &       $\mathbf{96.38\pm 0.72}$            &     $96.85\pm 0.84$              \\ \cline{2-5}
  &   1      &   1        &       $96.30\pm 1.12$            &     $\mathbf{97.36\pm 1.89}$              \\ \cline{2-5}
   & 1      &   2        &       $95.39\pm 1.26$            &     $96.44\pm 1.72$              \\ \cline{2-5}
   &  2      &   2        &       $95.68\pm 0.48$            &     $97.35\pm 1.01$              \\ \cline{2-5}
  &   2      &   3        &       $94.33\pm 1.48$            &     $96.53\pm 1.13$              \\ \cline{2-5}
   &  3      &   3        &       $94.86\pm 0.75$            &     $96.63\pm 1.09$              \\ \hline

\multirow{6}{6em}{MFCC-gram CNN}   &  0      &   1        &       $74.07\pm 1.93$            &     $61.11\pm 8.40$              \\ \cline{2-5}
  &   1      &   1        &       $86.11\pm 2.98$            &     $66.67\pm 3.74$              \\ \cline{2-5}
   & 1      &   2        &       $83.33\pm 3.34$            &     $84.26\pm 6.17$              \\ \cline{2-5}
   &  2      &   2        &       $85.19\pm 0.93$            &     $88.89\pm 0.53$              \\ \cline{2-5}
  &   2      &   3        &       $85.19\pm 1.85$            &     $74.07\pm 5.10$              \\ \cline{2-5}
   &  3      &   3        &       $87.04\pm 0.93$            &     $91.67\pm 2.98$              \\ \hline

\multirow{6}{6em}{Baseline Spectrogram Transformers}   &  0      &   1        &       $82.87\pm 1.48$            &     $68.73\pm 3.88$              \\ \cline{2-5}
  &   1      &   1        &       $82.84\pm 1.82$            &     $82.65\pm 2.25$              \\ \cline{2-5}
   & 1      &   2        &       $80.75\pm 2.22$            &     $77.18\pm 1.56$              \\ \cline{2-5}
   &  2      &   2        &       $79.02\pm 3.19$            &     $82.05\pm 2.57$              \\ \cline{2-5}
  &   2      &   3        &       $78.07\pm 2.78$            &     $74.67\pm 1.99$              \\ \cline{2-5}
   &  3      &   3        &       $78.73\pm 2.33$            &     $81.96\pm 1.97$              \\ \hline
\end{tabular}
\caption{The performance of the Transformers and the CNN is shown in Table~\ref{table:comparison}. The different lines show performance of the network according to the number of noise files added to the test files, both for patients and control.}\label{table:comparison}
\end{table}

We observe a significant improvement in performance for baseline MFCC-gram Transformers when compared to the MFCC-gram CNN. When including noise in test samples, the best performance is attained by baseline MFCC-gram Transformers where we add a single noise file to the control files and keep the patient files unchanged. When we compare without noise being added the best performance is attained by baseline MFCC-gram Transformers where we add a single noise file to both the patient and control files. We would like to point out though that the differences are rather small and baseline MFCC-gram Transformers performs well as long as some noise is added.

For the second experiment, we fix the amount of ward noise we insert to the training and test files to be a single noise file for both patient and control audio files. We vary the technique employed for unsupervised pretraining, attempting time alteration, channel alteration and noise alteration techniques as described in Section~\ref{subsec:unsupervisedpretraining}. We pretrained on the corpuses of NURC-Recife, C-Oral Brasil, SP 2010 and ALIP. Pretraining consisted of 5 epochs on the data of all those corpuses, splitting each file into $4$ seconds audio with a $1$ second window step. Finetuning on the respiratory insufficiency data was performed in $20$ epochs and repeated $10$ times so the results are averaged. We show the performance of each for both MFCC-gram Transformers and Spectrogram Transformers in Table~\ref{table:unsupervisedpretraining}.

\begin{table}[htb]
\begin{tabular}{|m{6em}|m{6em}|m{10em}|m{10em}|}
\hline
Model & Pretraining type & Accuracy (with noise in test samples) & Accuracy (without noise in test samples) \\ \hline
\multirow{6}{6em}{MFCC Transformers}   &  Baseline      &         $96.30\pm 1.12$            &     $97.36\pm 1.89$             \\ \cline{2-4}
  &   Time Alteration        &       $\mathbf{96.53\pm 0.71}$           &       $97.00\pm 1.55$          \\ \cline{2-4}
   & Channel Alteration        &      $96.15\pm 0.84$            &    $97.04\pm 1.52$               \\ \cline{2-4}
   &  Noise Alteration         &       $95.93\pm 0.66$            &   $98.21\pm 0.89$                \\ \cline{2-4}
   &  Time + Channel + Noise         &       $96.38\pm 1.24$            &   $\mathbf{98.54\pm 1.56}$                \\ \hline

\multirow{6}{6em}{Spectrogram Transformers}   &  Baseline      &        $82.84\pm 1.82$            &     $82.65\pm 2.25$             \\ \cline{2-4}
  &   Time Alteration        &        $80.99\pm 3.49$          &      $87.90\pm 2.75$            \\ \cline{2-4}
   & Channel Alteration        &       $82.41\pm 1.75$           &      $87.53\pm 2.25$             \\ \cline{2-4}
   &  Noise Alteration         &       $80.61\pm 1.32$            &   $86.08 \pm 2.70$                       \\ \cline{2-4}
   &  Time + Channel + Noise         &       $81.67\pm 1.51$            &   $86.93\pm 2.61$                \\ \hline

\end{tabular}
\caption{The performance of the Transformers network is compared when unsupervised pretraining is done. The different pretraining techniques are compared for MFCC-gram and Spectrogram Transformers. We fix the amount of noise insertion to be one noise file inserted at patient and control files.}
\label{table:unsupervisedpretraining}
\end{table}

We observe a small improvement (over the baseline) using time alteration when we test MFCC-gram Transformers including noise in the test files. We also observe an improvement using noise alteration when we test MFCC-gram Transformers without including noise in the test files. In principle, one could combine these techniques as they are independent ways of masking the input. We have done that by performing all three techniques at the same time as shown in the table. Note that the performance of Spectrogram Transformers increases even more robustly than that of MFCC-gram Transformers.

\section{Conclusion and Future work}

By employing a Transformers network to the dataset of respiratory insufficiency from COVID-19 detection created in the paper~\cite{spira2021}, we improved the performance of their CNN network from $87.04\%$ to $96.38\%$. Moreover, we found that MFCC and Spectrogram based Transformers improve their performance through unsupervised pretraining on a large amount of unlabeled data. 

Future work could involve augmenting the dataset with audios from patients of many more respiratory illnesses besides COVID-19. Moreover, we could ideally get audio from patients and control under similar conditions. Furthermore, one could attempt improving the performance of Spectrogram Transformers so that they match the performance of MFCC-gram Transformers. Moreover, we currently train our acoustic model in the single task of respiratory insufficiency detection. It would be interesting to extend our model for other tasks, creating the first acoustic model of spoken Brazilian Portuguese.

\section{Acknowledgement}

We would like to thank the LNCC for providing us with the computational resources required to do this work. All experiments were run in the LNCC servers. This work was supported by FAPESP grant number 2020/16543-7 (POSDOC) and project 06443-5 (SPIRA).  MF was supported by CNPq grant PQ 303609/2018-4.

\bibliographystyle{sbc}
\bibliography{refs}

\end{document}